%Paper: hep-th/9403022
%From: Sergei Skorik <sergei@phys1sun.usc.edu>
%Date: Thu, 3 Mar 1994 12:20:31 -0800

\input harvmac.tex
\Title{\vbox{\baselineskip12pt
\hbox{USC-94-002}}}
{\vbox{\centerline{Solution of the Thirring Model with Imaginary Mass}
\centerline{and Massless Scattering
}}}

\centerline{Hubert Saleur$^{\spadesuit *}$ and Sergei Skorik $^\dagger$ }
\vskip2pt
\centerline{$^\spadesuit$Department of Physics and Department of Mathematics}
\centerline{University of Southern California}
\centerline{Los Angeles CA 90089}
\centerline{$^\dagger$Department of Physics, University of Southern California}
\centerline{Los Angeles CA 90089}
\vskip2pt
\vskip.3in
The Thirring model with imaginary mass (or the sine-Gordon model with
imaginary coupling)  is deeply related to all
 the flows between minimal conformal theories. We solve this
model explicitely using
the Bethe ansatz. We find that there are Left and Right
moving massless excitations
with non trivial LR scattering. We compute the S matrix and recover the
result conjectured by Fendley et al.

\bigskip
\bigskip\bigskip
\noindent $^*$ Packard Fellow
\Date{3/94}

\newsec{Introduction.}

Massless integrable quantum field theories in 1+1 dimensions and the associated
 factorized "massless scattering" have been intensively studied recently
\ref\FS{P.Fendley, H.Saleur, "Massless integrable field theories and massless
scattering in 1+1 dimensions", preprint USC-93-022, and references therein.},
\ref\ZZ{Al.B.Zamolodchikov, Nucl. Phys. B358 (1991) 619, Nucl. Phys.
B366 (1991) 122 \semi
A.B.Zamolodchikov, Al.B.Zamolodchikov, Nucl Phys B379 (1992) 602.}.
The general context where such theories appear is when the perturbation of
 a conformal field theory by a relevant operator
causes it to flow to another conformal field theory. In the midst of the flow,
the model has massless excitations but is not scale-invariant.

In  ref. \ref\FSZ{P.Fendley, H.Saleur, Al.Zamolodchikov,
Int.J.Mod.Phys. A8 (1993) 5717, 5751.} it was shown that the sine-Gordon model
with imaginary potential (and its natural generalizations to higher
level and rank)
 plays a fundamental role in understanding flows between minimal
conformal field
theories, like the usual sine-Gordon model for massive perturbations.
 Exact (LL) and (LR) $S$  matrices
were found phenomenologically for this problem using the  Yang Baxter equation,
 unitarity and crossing-symmetry . The physical content of the model and the
 massless flows were then analyzed by thermodynamic Bethe ansatz (TBA).

The purpose of this letter is
to derive these $S$ matrices from the exact solution of the quantum
 field theory.   It is easy to check that the known equivalence between
sine-Gordon and Thirring extends to our case. We shall  therefore study
 the massive Thirring model (MT) with imaginary mass (IMT), which is easier
 to handle technically. We use the well known   coordinate Bethe
ansatz (BA) \ref\T{H.Bergknoff, H.Thacker, Phys. Rev. D19 (1979) 3666.}
, build the physical vacuum of the model and show explicitly
that there is no gap in the excitation spectrum.
 The model possesses non-trivial (LL) and (LR) scattering.
 The corresponding S-matrices are found from the BA equations
and  shown to be in agreement with the results of \FSZ.

\newsec{Structure of the model.}

The massive Thirring  (MT) model, defined by the  Hamiltonian
\eqn\Ham{H=\int dx [- i (\psi_1^+\partial_x\psi_1-\psi_2^+\partial_x\psi_2)
+m_0(\psi_1^+\psi_2+\psi_2^+\psi_1) + 2g\psi_1^+\psi_2^+\psi_2\psi_1],}
is a well-known example of an exactly soluble model.
Its quantization on a circle
of length $L$, first carried out in \T, leads to the Bethe ansatz equations
for the rapidities of $N$ pseudoparticles:
\eqn\bethe{\eqalign{e^{- i  L p(\beta_i)}&=\prod_{j=1}^N
{\sinh{1\over 2}(2 i \mu - \beta_i +\beta_j)\over \sinh{1\over 2}
(2 i \mu + \beta_i -\beta_j)}, \cr
-p(\beta_i)L &=\sum_j^N \Phi(\beta_i-\beta_j)+2\pi I_i . \cr}}
Here $p(\beta)$ is the momentum of a particle, $\Phi(\beta_i-\beta_j)$ is
the  two-particle bare phase shift,
\eqn\phase{\Phi(\beta)=- i \log{\sinh{1\over 2}(2 i \mu -\beta) \over
\sinh{1\over 2}(2 i \mu +\beta)} , \qquad \cot\mu=-{1\over 2}g,      }
and $I_i$ is a set of $N$ integer numbers. The momentum and energy $h(\beta)$
of a particle lie on the hyperbola $h^2-p^2=m_0^2$, which is parametrized
in terms of rapidities $\alpha$:
$$p=m_0\sinh\alpha, \qquad h=m_0\cosh\alpha.$$
Suppose now that the mass becomes purely imaginary, $m_0= i  m$.
The hyperbola gets rotated,
$p^2-h^2=m^2$, and momentum exchanges its role with energy. For the energy
and momentum to be real, the rapidity should now be $\alpha=- i {\pi\over 2}
+\beta$, where $\beta$ belongs to
$\{$Im$\beta=0\}\bigcup\{$Im$\beta=\pi\}$.
The new parametrization in terms of $\beta$ is
\eqn\moment{p=m\cosh\beta, \qquad h=m\sinh\beta.}
 This affects the solutions of equations \bethe\ and modifies drastically the
structure of the ground state and excitations
of the Thirring model. Note that the presence of an imaginary mass in the
Hamiltonian \Ham\ makes it non-hermitian.
However it is probable that the spectrum
is real. Indeed complex values of $\beta$ are
 strongly restricted by the Bethe equations. The
latter allow only string type solutions with a
definite pattern of complex $\beta$'s, for which the  total energy is real.

In the thermodynamic limit, equations \bethe\ become coupled integral
equations for the set of densities $\rho_j(\beta)$. Let us discuss
first the ground state of the model. The densities of the ground state
 distribution $\rho_{\pm}$ satisfy the
following system of integral equations (dots denote  derivative
with respect to $\beta$):
\eqn\dens{\eqalign{&\dot{p}_+(\beta)=
    -\int_{-\infty}^B\dot{\Phi}_{1,1}(\beta-\alpha)\rho_+(\alpha)d\alpha -
\int_{-B}^{\infty}\dot{\Phi}_{-1,1}(\beta-\alpha)\rho_-(\alpha)d\alpha
 - 2\pi\rho_+(\beta) \cr
    &\dot{p}_-(\beta)= -
\int_{-\infty}^B\dot{\Phi}_{-1,1}(\beta-\alpha)\rho_+(\alpha)d\alpha -
\int_{-B}^{\infty}\dot{\Phi}_{1,1}(\beta-\alpha)\rho_-(\alpha)d\alpha -
                                       2\pi\rho_-(\beta)          \cr}}
where
$\Phi_{-1,1}(\beta+ i \pi)=\Phi_{1,1}(\beta)\equiv\Phi(\beta)$ (see \phase).
The subscript $+$ (resp. $-$) means that the corresponding function
 is defined on the axis Im$\beta=0$ (resp. Im$\beta=\pi$). $B$ is a cut-off
 such that $\rho_+=0$ for $\beta>B$ and $\rho_-=0$ for $\beta<-B$.
Equations \dens\ define a one-parameter family of solutions $\rho_{\pm}(
\beta, B)$. The ground state will be found in principle by requiring that
$\rho_{\pm}\geq 0$ and that the total energy,
\eqn\energy{E_0=\int_{-B}^{\infty}h_-\rho_- + \int_{-\infty}^B h_+\rho_+,}
is minimal. One also has to check that this ground state is stable,
ie that all excitations have positive energy. Note that the presence of
pseudoparticles with positive bare energy in the ground state cannot be
ruled out at this stage. The filling of such states
brings some positive contribution to $E_0$, but it affects also
the distribution
of all the other pseudoparticles through the Bethe equations
 \bethe, an effect which can lower the total energy.

Let us comment on the sign of the last term at the right hand side of \dens.
This sign depends on whether the integers $I_i$
 in \bethe\ are an increasing or decreasing function of the bare rapidity.
In other words it indicates the signature of the derivative
$\dot y(\beta)$ where $p(\beta)L + \sum_j \Phi(\beta - \beta_j)\equiv y(\beta)$
 (assuming this is $\beta$-independent). In the simplest situation, when
the signs of $\dot{p}(\beta)$ and $\sum_j\dot{\Phi}(\beta - \beta_j)$
coincide for all
$\beta$ one can make a rigorous conclusion about the signature of $\dot y$.
This was the case in the usual Thirring model. Otherwise, the evaluation
of sign$(\dot y)$ becomes more delicate  because of the infinite
summation
over the so far unknown positions $\beta_j$, the parameters $L$, $N$
and the momentum cutoff which should be sent to infinity. Close to the
 free point one  expects however  that the sign of $\dot{y}$ is  the sign
of $\dot{p}$ for the rapidities of filled states; this is how we choose the
sign in equations \dens\ .  Our choice gives
  a reasonable solution of \dens\ for which  $\rho_{\pm}>0$ and $B=\infty$.
Other choices seem to lead
to negative densities,  but it is difficult to argue
 for finite $B$ except close to the free point.

It is known that in order to solve the MT
model appropriate regularization is required \T. To eliminate the
divergences we
introduce an analytic momentum cutoff similar
to the one used in the $U(1)$-symmetric Thirring model with two isospin
degrees of freedom
\ref\W{G.Japaridze, A.Nersesyan, P.Wiegmann, Nucl. Phys. B230 (1984) 511.}
 and in the XXZ lattice model
\ref\YY{C.N.Yang, C.P.Yang, Phys. Rev. 150 (1966) 327.},
\ref\TS{M.Takahashi, M.Suzuki, Progr. Theor. Phys 48 (1972) 2187.}:
\eqn\cutoff{\eqalign{p_+(\beta)&=
 i m\log{\cosh{1\over 2}(\beta-f- i \mu) \over
\cosh{1\over 2}(\beta-f+ i \mu)} +  i m\log
{\sinh{1\over 2}(\beta+f- i \mu) \over\sinh{1\over 2}(\beta+f+ i \mu)}\cr
&\equiv G(\beta-f)+G(\beta+f+i\pi), \cr
\dot{p}_+(\beta)&=m\sin\mu\left({1\over\cosh(\beta - f)+\cos\mu} -
             {1\over\cosh(\beta + f)-\cos\mu}\right),\cr}}
and $\dot{p}_-(\beta)=\dot{p}_+(\beta+ i \pi)$.
The peculiarity of the cutoff \cutoff\ is an unusual  shift
of argument by $i\pi$ between the first and second terms.
 Upon the cutoff removal,
$f\rightarrow\infty$, we recover our momentum of pseudoparticles:
$p_+\rightarrow 4m\sin\mu\exp(-f)\cosh\beta$, $\quad
p_-\rightarrow -4m\sin\mu\exp(-f)\cosh\beta$.

In principle,
the system of integral equations \dens\ cannot be solved analytically
for arbitrary $B$.  Using the symmetry $\rho_+(\beta)=\rho_-(-\beta)$,
that follows from  $\dot{p}_-(-\beta)=\dot{p}_+(\beta)$,
the system \dens\ can be rewritten as a single Fredholm integral equation:
\eqn\fred{\dot{p}_+(\beta)
=-\int_{-\infty}^B \left(\dot{\Phi}_{1,1}(\beta-\alpha) +
\dot{\Phi}_{-1,1}(\beta+\alpha)\right) \rho_+(\alpha)d\alpha -
                                                   2\pi\rho_+(\beta).    }
Equations of such a form
have previously appeared in related problems, for instance
the $U(1)$-symmetric Thirring
model in a magnetic field
(WC sector) \W, \YY. An approximate solution can be obtained if one treats
the second kernel in \fred\ as a small perturbation to the first one and then
makes use of the Wiener-Hopf method.

We show that in the repulsive regime
($0<\mu<\pi/2$)  the solution of \dens\ with
$B=\infty$ is consistent, and we conjecture that
 it represents the ground state of the model at least
for $0<\mu<\pi/4$ (strong repulsive regime). Our main argument is
that such a solution leads to  reasonable density distributions,
to positive excitation
energies over the physical vacuum, and to the correct scattering matrices
(see section 3).
 Solving equations \dens\ with $B=\infty$ by Fourier transformation
and using
$$ \int_{-\infty}^{\infty}\dot{\Phi}_{1,1}(\beta)e^{ i  k\beta}d\beta =
2\pi{\sinh(\pi-2\mu)k \over \sinh\pi k}, \qquad
 \int_{-\infty}^{\infty}\dot{\Phi}_{-1,1}(\beta)e^{ i  k\beta}d\beta =
-2\pi{\sinh 2\mu k \over \sinh\pi k},$$
$$\int_{-\infty}^{\infty}{\dot{p}_{+}(\beta)\over 2\pi}e^{ i  k\beta}d\beta =
   m e^{ i  kf}{\sinh\mu k \over \sinh\pi k} - m e^{- i  kf}
{\sinh(\pi-\mu)k\over\sinh\pi k},                                      $$
we obtain:
\eqn\eqI{ {\hat{\rho}_+(k) \choose \hat{\rho}_-(k)} = {m\over 2\cosh\mu k}
           {e^{- i  kf}   \choose  e^{ i  kf}  }.  }
The inverse Fourier transform $\rho(\beta)=\int\hat{\rho}(k)\exp(- i  k
\beta)dk/2\pi$ yields:
\eqn\eqII{\rho_+ = {m\over 4\mu\cosh{\pi\over 2\mu}(\beta + f)},
\qquad         \rho_- = {m\over 4\mu\cosh{\pi\over 2\mu}(\beta - f)}.  }
Upon  cutoff removal,
$$\rho_{\pm}(\beta)\rightarrow {m\over 2\mu} e^{-{f\pi\over 2\mu}}
              e^{\mp{\beta\pi\over 2\mu}}, \qquad f\rightarrow\infty .   $$
The density $\rho_+ (\rho_-)$ decrease exponentially for those $\beta$, where
$h(\beta)\gg m$, i.e. $\beta\rightarrow +\infty$ $(\beta\rightarrow -\infty)$.
On the other side of the deep ultraviolet region,
$\beta\rightarrow -\infty$ $(\beta\rightarrow +\infty)$,
where $h(\beta)\ll -m$, the functions \eqII\ have the same asymptotic
behaviour,
as the density in the usual MT model \W. This is expected since for
 $\beta\rightarrow -\infty$  $(\beta\rightarrow +\infty)$ we can neglect
the contribution from the pseudoparticles located on Im$\beta=\pi$
 (Im$\beta=0$) line in eqs. \bethe.
When $\mu\rightarrow 0$, $\quad\rho_{\pm}(\beta)\sim H(\mp\beta)$, where
$H(x)$ is the Heaviside step function.

To obtain the energy spectrum of excitations at $T=0$ we use the TBA method.
Variation of densities of pseudoparticles over the physical vacuum
leads to the system of integral equations \W:
\eqn\TBA{\eqalign{& h_j=\varepsilon_j^+ + A_{jk}\ast\sigma_k\varepsilon^-_k \cr
        & p_j=\pi_j^+ + A_{jk}\ast\sigma_k\pi^-_k , \cr} }
where $\ast$ denotes convolution, $\varepsilon_j^{\pm}$ and $\pi_j^{\pm}$
are the energies and momenta of different excitations labelled by $j$
($\varepsilon^+>0$ and $\varepsilon^-<0$ by definition), and
$$A_{jk}=
-\dot{\Phi}_{jk}/2\pi +\sigma_j\delta_{jk}\delta(\beta),$$
where $\sigma_j=-1$ for
the holes ($j=\pm 1$) and $\sigma_j=1$ for  $|j|\geq 2$.  For technical
simplicity we have  restricted
the coupling parameter to rational values $\mu=\pi/s$, $s$ being
an integer. In this case the allowed strings are those of  length
from 2 to $s-1$, centered around the axis $\hbox{Im}\beta=0,\pi$ which we
denote
by $\pm j$ (strings and antistrings in the usual denomination). To see
this one follows the usual argument,  taking the
modulo of the both sides of \bethe\ \TS. When there is a real $\beta_i$
in the left hand side of \bethe, the modulo of both sides is easily seen to be
1. When the rapidity of one of the particles of the string is plugged into the
L.H.S., one of the terms in the R.H.S. becomes equal to zero or infinity.
The L.H.S. vanishes or blows up as well in the limit $L\rightarrow\infty$,
 depending on the sign of the imaginary part of the momentum.  The
latter is however a cumbersome expression depending on the cut-off $f$; this is
 not usually the case. If one takes
the limit $f\rightarrow\infty$ at first, one reproduces the expression for
the momentum \moment, from which the result follows straighforwardly.

We choose the regularization of the bare energy to be
 $h_j(\beta)=\dot{p}_j(\beta)$,
see \cutoff\ (another regularization, such as $h_j(\beta)=G(\beta-f)-G(\beta+f+
i\pi)$ could lead to different mass renormalization, which in our case is not
important).

For the holes in the ground state \eqII, solution of \TBA\ with
$\varepsilon^+_{\pm 1}=0$ yields in the limit $f\rightarrow\infty$:
$$\varepsilon^-_{\pm 1}(\beta)=- {m\pi\over \mu} e^{-{f\pi\over 2\mu}}
              e^{\mp{\beta\pi\over 2\mu}},    $$
$$\pi^-_{\pm 1}=\mp {2\mu\over\pi}\varepsilon^-_{\pm 1}. $$
These excitations are
massless left and right-moving particles which we call by analogy  "solitons".

For the strings of length $|j|=2,...,s-2$ we obtain
$\varepsilon^+_j=\pi^+_j=0$,
where we used
$$\hat{A}_{1,j}(k)=-2{\cosh\mu k \sinh(\pi-j\mu)k  \over  \sinh\pi k}, \qquad
  \hat{A}_{-1,j}(k)=2{\cosh\mu k \sinh j\mu k  \over  \sinh\pi k}. $$
For the longest ($s-1$) string the calculation yields
$\varepsilon^+_{\pm (s-1)}=|\varepsilon^-_{\mp 1}|$. Such a string with the
center
located on the line Im$\beta=0$ (Im$\beta=\pi$) behaves as a hole with the
same rapidity located on the line Im$\beta=\pi$ (Im$\beta=0$).

There is one more peculiar excitation in the IMT model
in thermodynamic limit. This is a 1-string, located on the imaginary rapidity
 axis
exactly in between the two filled vacuum lines
(see fig.2), with  $\beta= i \pi/2$.
To check that $\beta=i\pi/2$ is indeed a solution of \bethe\
in the thermodynamic limit we take the modulo of both sides of \bethe,
where the momentum is given by \cutoff, and set
$L,f,N\rightarrow\infty$.
  If one sets $\beta_i=x+ i \pi/2$ in the left hand side and
combines the terms with $\beta_j$ and $\beta_{j'}=2x-\beta_j+ i \pi$ in the
R.H.S. of \bethe, the product of terms in the R.H.S. gives 1. This can be
seen also by symmetry of the picture, depicted in fig.2. In the L.H.S., the
momentum in the exponential is proportional to $m\exp(-f)$. In order to keep
the physical (dressed) mass finite, one should introduce the cut-off dependence
in the bare mass, $m\sim\exp({f\pi\over 2\mu})$. Then, in the limit
$L\rightarrow\infty$ the modulo of the L.H.S. is 1 iff $x=0$,
independently from $f$.
 If one chooses $\beta_i$ real in L.H.S. of \bethe, then
 $\beta_j=i\pi/2$ will appear in the R.H.S., making its modulus to be
different from 1. So it seems $\beta=i\pi/2$ is not allowed.  However,
observe that multiplying this equation by  \bethe\
with $\beta_{i'}=-\beta_i+ i\pi$ we obtain 1 in the R.H.S., too.
This suggests that the Bethe equations describing such 1-string should
be combined into pairs according to the symmetry shown on fig.2.
The possible physical explanation is that the vacuum of the IMT
model consists of a condensate of {\bf pairs} of L and R moving particles with
opposite momenta and equal energy. The appropriate Bethe equations
 would then represent the periodic boundary conditions, written
for the wave function when one moves such a pair
through the system. The formation of pairs appears actually necessary to
 conserve parity, which is not broken in our problem.

 The physical (dressed) energy and momentum of this new excitation,
calculated by
means of \TBA\ vanish.
$$ \varepsilon^+_0=\pi^+_0 = 0, $$
where we used the phase shifts \phase\ given by
\eqn\phass{{\dot{\Phi}_{1,0}\choose \dot{\Phi}_{-1,0}} = -2\pi{\sinh 2\mu
k\over
\sinh \pi k}{e^{-{\pi k\over 2}}\choose e^{{\pi k\over 2}}} ,}
and
$$ h_0 (k) = 4\pi{\sinh\mu k\over \sinh\pi k}\cosh({k\pi\over 2}+
 i  k f).        $$
This result still holds formally for excitations $\beta=x+i{\pi\over 2}$, $x$
real.

We conclude this section by  remarks on the structure of the IMT model
in the attractive regime and in the vicinity of the free fermion point in
the repulsive
regime. When one goes to the attractive regime ($\mu>\pi/2$), the
 two-particle bare phase shifts \phase\ change sign: $\Phi(\beta)\rightarrow
-\Phi(\beta)$ when $\mu\rightarrow\pi-\mu$.
 The solution
of \dens\ with $B=\infty$ is no longer valid since it leads
to negative densities. Therefore, $B$ must be finite.
It is natural to suppose that a
 phase transition happens in the system at $\mu=\pi/2$
or at $\mu=\Delta<\pi/2$, where $\Delta$ takes one of the possible values
$\pi/3$ or $\pi/4$. Such a phase transition would probably  involve level
 crossing,
like for  the MT model in the repulsive regime and appropriate cut off,
 see \ref\xXx{V.Korepin, Comm.Math.Phys. 76 (1980) 165 \semi
Itoyama, Moxhay, Phys. Rev. Lett. 65 (1990) 2102}.
%It can be analyzed
%by comparing the energies of different ground state configurations as
%it was done for the lattice Potts model \ref\Potts{Albertini, McCoy, Perk,
%Phys. Lett. 135A (1989) 159; 139A (1989) 204}.

In the limit $\mu\rightarrow\pi/2$ there are
two different solutions to \dens, obtained for $B=\infty$ and $B=0$.
The first one is given by
\eqII, and the second  is $2\pi\rho_{\pm}=-\dot{p}_{\pm}$
(defined on the appropriate half-line). The ground  state
energies \energy\ for  these two solutions turn out to be identical,
 which suggests that the phase transition takes place exactly at $\mu=\pi/2$,
 and that the
 free point is singular, as anticipated in \FSZ.   A rigorous mathematical
 analysis of the integral equation \fred\ or a detailed numerical solution
(which is delicate because of cut off effects) could shed some light on this
 question.

As in the real mass case \xXx\  we expect that the physical results depend
on the type of
cut-off. It is possible that
a very  different behaviour  may emerge for instance if one uses a sharp
cut-off such as
$p_\pm(\beta)=0$ if $|\beta|>\Lambda$.

\newsec{Scattering of excitations. }

 In order to compute  elastic S-matrices we adopt the method developed in
\ref\K{V.Korepin, Theor. Math. Phys. 41 (1979) 953.}.
Let us consider holes at first.
By definition, the phase shift for scattering of two holes is given by

\eqn\eqi{\delta_h(\beta_1-\beta_2)\equiv{1\over i }\log S = \varphi_1 -
                                                                \varphi_2,}
where $\varphi_1$ is the phase gained by a hole when  going around
the system and $\varphi_2$ the same phase but in the presence of another hole.
The $\varphi_{1,2}$
are composed of a sum of two-particle bare phase shifts; for example
$$ \varphi_2 = Lp_{\pm}(\beta_1) + \sum_j \Phi_{1,1}(\beta_1-\beta_j)+ \sum_j
                                    \Phi_{-1,1}(\beta_1-\beta_j).         $$
Here the sum is taken over all the solutions $\beta_j$
of the Bethe equations \bethe\ with two holes at positions $\beta_1$ and
$\beta_2$. The result of subtraction $\varphi_1 - \varphi_2$
is proportional to  the
backflow function of vacuum: $\delta_h = 2\pi F(\beta_2|\beta_1)$.
The latter is defined using the difference of the two  solutions of
the Bethe equations obtained  with and without a hole  at $\beta_0$,
$F(\beta_0|\beta)\equiv(\beta-\tilde\beta)L\rho(\beta)$. In our case we have
two  backflow functions $F_+$, $F_{-}$, describing a backflow on
Im$\beta=0$ and Im$\beta=\pi$ lines respectively. One can show using \bethe\
that $F_{\pm}$ satisfy the following system of integral equations:
\eqn\eqii{\eqalign{&\Phi_{1,1}(\beta-\beta_0)=\dot{\Phi}_{1,1}\ast F_+ +
                              \dot{\Phi}_{-1,1}\ast F_- + 2\pi F_+ \cr
                    &\Phi_{-1,1}(\beta-\beta_0)=\dot{\Phi}_{1,1}\ast F_- +
                              \dot{\Phi}_{-1,1}\ast F_+ + 2\pi F_- \cr}}
Equations \eqii\ describe the backflow caused by a hole at $\beta=\beta_0$
on the real axis; for a hole on Im$\beta=\pi$ axis simply substitute
 $F_+\leftrightarrow F_-$ in eqs \eqii.
Taking a derivative with respect to $\beta$ and applying the Fourier transform
to both sides of \eqii\ we arrive at the following solution (in Fourier space):
\eqn\eqiii{{\dot{F}_+^h  \choose \dot{F}_-^h} =
 {e^{ i  k \beta_0}\over 2\cosh\mu k \sinh(\pi-2\mu)k}
 {\sinh(\pi-3\mu)k\choose -\sinh\mu k},}
$F_+$ represents scattering of two holes on the same line ( ie LL or RR
scattering) and $F_-$ scattering of two holes on different lines
( ie LR or RL scattering). From \eqi\ we obtain:
\eqn\eqt{{1\over i }{d\over d\theta}\log A_{LL}(\theta)=
\int_{-\infty}^{+\infty}
e^{- i  k\theta}{\sinh(\pi-3\mu)k\over 2\cosh\mu k \sinh(\pi-2\mu)k}dk,}
\eqn\eqtt{{1\over i }{d\over d\theta}\log A_{RL}(\theta)=
 - \int_{-\infty}^{+\infty}
e^{- i  k\theta}{\sinh\mu k\over 2\cosh\mu k \sinh(\pi-2\mu)k}dk,}
where $\theta\equiv \beta_1 - \beta_2$. Note that this method
does not fix the constant
normalization factor in the matrix elements $A_{LL}, A_{RL}$,
which should be fixed by other constraints (e.g. unitarity). Eventually,
we obtain:
\eqn\eqj{ A_{LL}(\theta)=\exp{i\over 2}
\int_{-\infty}^{+\infty}
\sin k\theta{\sinh(\pi-3\mu)k\over \cosh\mu k \sinh(\pi-2\mu)k}{dk\over k},}
\eqn\eqjj{ A_{RL}(\theta)=
  i \exp-{i\over 2} \int_{-\infty}^{+\infty}
\sin k\theta{\sinh\mu k\over \cosh\mu k \sinh(\pi-2\mu)k}{dk\over k}.}
Expressions \eqj\ and \eqjj\ are
in agreement with results of \FSZ, which enables us to identify coupling
constants in imaginary mass MT model with those of imaginary potential SG
model:
$\mu=\pi - \beta_{SG}^2/8$.
 Note that the integrand in \eqj\ and \eqjj\ blows up when
$\mu \rightarrow \pi/2$ (free fermions).
The LL-scattering becomes trivial when $\mu \rightarrow \pi/3$, whereas
for the LR-scattering  this happens when $\mu\rightarrow 0$.
The L and R systems are decoupled at $\mu=0$ and the model
not only has massless excitation spectrum, but also is scale invariant.

In the same manner one can calculate the backflow functions for strings.
We will need them to get the rest of the matrix elements for the
soliton-antisoliton S-matrix. For the j-string (centered around the real axis)
one has a system, similar to
\eqii:
\eqn\eqiis{\eqalign{&-\Phi_{1,j}(\beta-\beta_0)=\dot{\Phi}_{1,1}\ast F_+ +
                              \dot{\Phi}_{-1,1}\ast F_- + 2\pi F_+ \cr
                    &-\Phi_{-1,j}(\beta-\beta_0)=\dot{\Phi}_{1,1}\ast F_- +
                              \dot{\Phi}_{-1,1}\ast F_+ + 2\pi F_- \cr}}
The solution to these equations for $j<s-1$ is:
\eqn\eqiiis{\dot{F}_+^s = -{\sinh(\pi-(j+1)\mu)k\over\sinh(\pi-2\mu)k}
e^{ i  k \beta_0}, \qquad
\dot{F}_-^s = {\sinh(j-1)\mu k\over\sinh(\pi-2\mu)k}e^{ i  k \beta_0}}
and for  $j=s-1$
\eqn\eqiiiss{\dot{F}_+^s = -{\sinh\mu k\over 2\cosh\mu k \sinh(\pi-2\mu)k}
e^{ i  k \beta_0}, \qquad
\dot{F}_-^s = {\sinh(\pi-3\mu)k\over 2\cosh\mu k \sinh(\pi-2\mu)k}
e^{ i  k \beta_0},}
which is the same as the solution for the hole on Im$\beta=\pi$ axis.

To address the question of soliton antisoliton scattering it is useful to
 determine the charge of excitations, i.e. the
eigenvalues of the operator $N=\int\psi_1^+\psi_1 + \psi_2^+\psi_2$.
Setting the charge of the vacuum to zero, one has
\eqn\eqz{Q_h=-1+\int_{-\infty}^{+\infty}(\dot{F}_+^h + \dot{F}_-^h)d\beta=
              -{\pi\over 2(\pi-2\mu)}}
for a hole and
\eqn\eqzz{Q_{str}=j+\int_{-\infty}^{+\infty}(\dot{F}_+^s + \dot{F}_-^s)d\beta=
              {\pi(j-1)\over \pi-2\mu}}
for the j-string, where $j<s-1$ and \eqiii, \eqiiis\ were used. The simplest
configuration of vanishing charge can be composed
of two holes and a 2-string (see fig.1). It reproduces
LL (RR) soliton-antisoliton S-matrix. Indeed, one can see
using the  definition \eqi\ that the additional phase which the hole
($\beta_1$)  acquires in the presence of the
 string ($(\beta_1+\beta_2)/2$) and another hole ($\beta_2$)
consists of the sum of two phase shifts of a hole-hole and string-hole
scattering:
$\delta_{hsh}(\theta)=\delta_h(\theta)+\delta_{hs}(\theta/2)$. The hole-hole
scattering phase was calculated previously (see \eqj). As for the string-hole
phase shift, it is given by the solution of equations \eqiis\ for $j=2$ and
$\beta_0=(\beta_1+\beta_2)/2$, $F_+$ and $F_-$ being related to the
transmission
and reflection amplitudes. Taking the inverse Fourier transform of \eqiiis\ we
obtain after some manipulations:
\eqn\eqsa{U^+_{LL}=-{\sinh{\pi\over4}({\theta\over\pi-2\mu}+
{2 i \mu\over\pi-2\mu}
)\over \sinh{\pi\over4}({\theta\over\pi-2\mu}-{2 i \mu\over\pi-2\mu})},
\qquad
U^-_{LL}=-{\cosh{\pi\over4}({\theta\over\pi-2\mu}+{2 i \mu\over\pi-2\mu}
)\over \cosh{\pi\over4}({\theta\over\pi-2\mu}-{2 i \mu\over\pi-2\mu})}.}
As in the usual Thirring model, solutions of the Bethe ansatz equations
correspond
to  linear combinations of multiparticle states for which the
scattering  is diagonal. In the usual basis the matrix
elements would be $A_{LL}$ (describes $ss\rightarrow ss$ process),
 $B_{LL}$ (describes $sa\rightarrow sa$ process), and $C_{LL}$ (describes $sa
\rightarrow as$ process), where
$B_{LL}\pm C_{LL}=A_{LL}U_{LL}^{\pm}$.

To describe the LR (RL) scattering  we introduce two holes and a 1-string
at $(\beta_1+\beta_2+ i \pi)/2$, formally without the constraint that the
sum of rapidities vanishes (as illustrated on fig.2 for $\beta_1=-\beta_2)$.
The backflow due to our new 1-string can be obtained from the system \eqiis.
The result is
\eqn\eqpi{ {\dot{F}_+  \choose  \dot{F}_- } = e^{ i  k (\beta_1+\beta_2)/2}
{\sinh\mu k\over\sinh(\pi-2\mu)k}
           {e^{(\mu-{\pi\over 2})k}   \choose  e^{({\pi\over 2}-\mu)k}  }.  }
The charge of this 1-string can be evaluated by formula \eqzz\ to be:
\eqn\eqzzz{Q_{ i \pi/2} =  {\pi\over \pi - 2\mu}        ,}
so that the total charge of our state is $Q_{ i \pi/2}+2Q_h=0$.
As in  the case of LL-scattering,  the functions \eqpi\ lead to the string-hole
phase shifts which are related to the LR transmission-reflection amplitudes. We
get:
\eqn\eqsa{U^-_{RL}={\sinh{\pi\over4}\bigl({\theta\over\pi-2\mu}+
{2 i \mu\over\pi-2\mu} -  i \bigr)
\over \sinh{\pi\over4}
\bigl({\theta\over\pi-2\mu}-{2 i \mu\over\pi-2\mu}- i \bigr)}
\qquad
U^+_{RL}={\cosh{\pi\over4}
\bigl({\theta\over\pi-2\mu}+{2 i \mu\over\pi-2\mu}- i \bigr)
\over \cosh{\pi\over4}\bigl({\theta\over\pi-2\mu}-{2 i \mu\over\pi-2\mu}
- i \bigr)}.}
The hole-hole scattering is given by \eqjj. Putting everything together, we
obtain the LR soliton-antisoliton S-matrix elements $A_{RL}$,
$B_{RL}$ and $C_{RL}$, where $B_{RL}\pm C_{RL}=A_{RL}U_{RL}^{\mp}$.
This S-matrix is identical to the one found in  \FSZ\  by solving the
YBE.
One should remember however the constraint on the position of the 1-string,
obtained at the end of section 2, that imposes to consider only
the scattering of pairs of L and R moving particles with opposite momenta.
Moreover
note that due to the presence of purely exponential factors in the Fourier
images $F_{\pm}(k)$, the functions $F_{\pm}(\beta)$  have imaginary parts
(however, the combination $F_+ + F_-$ is a real function).
This means that  particles of the vacuum are pushed off the
Im$\beta=0$ and Im$\beta=\pi$ lines in the complex rapidity plane towards
the particle at $(\beta_1+\beta_2+ i \pi)/2$. So it may be that the complete
 scattering
theory is more complicated than the one of \FSZ. Probably a numerical study of
the
appropriate inhomogeneous 6-vertex model could shed light on this.

 Note that there are misprints in the formulas (2.7) of \FSZ, page 5757.
The correct form of these equations is given in the preprint version of \FSZ.

\bigskip

\noindent{\bf Acknowledgments}: This research  was supported the National
Young Investigator program (NSF-PHY-9357207),
 by DOE  grant No. DE-FG03-84ER40168 and the Packard foundation.

\listrefs

\centerline{\bf Figure captions}
\smallskip
\noindent Figure 1: a pair of holes and a 2-string centered on the real axis
describe a pair of L moving "soliton" and "antisoliton".
\smallskip
\noindent Figure 2: a pair of holes with opposite rapidities on the real axis
 and the
axis $\hbox{Im}\beta=\pi$ together with a string at $\beta=i\pi/2$ describe a
 pair of
L and R  moving "soliton" and "antisoliton".

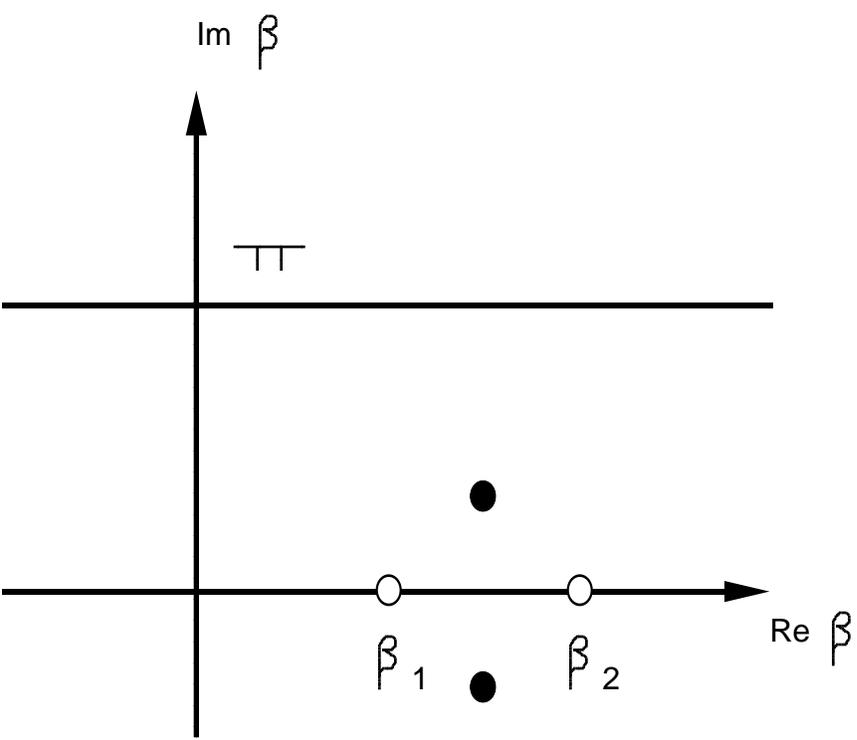
\bye